\documentclass[12pt]{article}
\usepackage{epsfig}
\begin{document}
\begin{center}
{\Large {\bf Torsion instead of Technicolor}

\vskip-40mm \rightline{\small ITEP-LAT/2010-03 } \vskip 30mm

{%\baselineskip=16pt
\vspace{1cm}
{ M.A.~Zubkov\footnote{e-mail: zubkov@itep.ru } }\\
\vspace{.5cm} {\it  ITEP, B.Cheremushkinskaya 25, Moscow, 117259, Russia } }}
\end{center}

\begin{abstract}
We consider the model, which contains a nonminimal coupling of Dirac spinors to
torsion. Due to the action for torsion that breaks parity the left - right
asymmetry of the spinors appears. This construction is used in order to provide
dynamical Electroweak symmetry breaking. Namely, we arrange all Standard Model
fermions in the left - handed spinors. The additional technifermions are
arranged in right - handed spinors. Due to the interaction with torsion the
technifermions are condensed and, therefore, cause appearance of the gauge
boson masses. In order to provide all fermions with masses we consider two
possibilities. The first one is related to an additional coupling of a real
massive scalar field to the considered spinors. The second possibility is to
introduce the explicit mass term for the mentioned Dirac spinors composed of
the Standard Model fermions and the technifermions.
\end{abstract}

\section{Introduction}

It is well - known that quantum gravity in the first order formalism with
either Palatini action or Holst action leads to a four - fermion interaction
between spinor fields coupled in a minimal way to torsion \cite{Rovelli}. The
given four - fermion interaction may lead to condensation of the fermions under
certain conditions. A condensation of this kind has been considered in some
cosmological models as a source of dark energy
\cite{Xue,Alexander1,Alexander2}.

In the present paper we suggest to consider torsion coupled in a nonminimal way
to fermion fields \cite{Shapiro,Shapiro_,Shapiro__,Shapiro___}  as a source of
Dynamical Electroweak symmetry breaking (DEWSB). The basic idea is to assume
that Poincare quantum gravity has two different scales. The first one is
related to metric field and the Riemannian connection compatible with metric.
This scale is supposed to be somewhere around Planck mass and does not affect
the low energy physics we deal with. The second scale is related to torsion
degrees of freedom. This second scale is assumed to be one or two orders of
magnitude over the TeV scale. That's why we expect torsion may play an
important role in a new physics expected at the TeV scale.

Probably, the most popular scheme of DEWSB is related to the Technicolor (TC)
theory \cite{Technicolor0,Technicolor,Technicolor_,Technicolor__}. This theory
contains an additional set of fermions that interact with each other via the
Technicolor gauge bosons. This interaction is of the attraction nature and,
therefore, in an analogy with the BCS superconductor theory may lead to
formation of the condensate composed of fermions. This condensate, in turn,
breaks Electroweak symmetry down to Electromagnetism. The effective Nambu -
Jona - Lasinio model (NJL) of chiral symmetry breaking \cite{ENJL} in TC
contains four fermion interactions similar to those obtained in the theory of
torsion coupled to fermions. This prompts us that the latter can be used, in
principle, instead of technicolor. It is worth mentioning that TC theory itself
suffers from problems related to different mechanisms of fermion mass
generation. Usually suggested Extended Technicolor (ETC) interactions
\cite{ExtendedTechnicolor0,ExtendedTechnicolor,ExtendedTechnicolor_,ExtendedTechnicolor__,ExtendedTechnicolor___,ExtendedTechnicolor____}
do not pass precision Electroweak tests due to FCNC and contributions to
Electroweak polarization operators. The so-called walking technicolor
\cite{walking,minimal_walking,minimal_walking_,minimal_walking__} improves the
situation but does not allow to generate $t$ - quark mass in an appropriate
way. Certain models in bosonic technicolor \cite{BosonicTC,BosonicTC_} (that
use the exchange by scalar particles instead of ETC gauge bosons)
 allow to generate fermion masses without
the problems specific for ETC (see, for example,
\cite{Kagan,Kagan_,Kagan__,Dobrescu_Kagan,Dobrescu_Kagan_,Dobrescu_Kagan__,Z2010_2}).
But these models do not solve the Hierarchy problem as the mass terms for the
techniscalars receive quadratically divergent contributions from loop
corrections.

It is worth mentioning that the torsion field coupled in a minimal way to
fermions cannot alone produce the correct DEWSB as it is coupled to all
fermions in an equal way. Suppose that the additional fermions aimed to form a
condensate are introduced. Then the condensation of these fields may occur only
together with the condensation of the Standard Model (SM) fermions. In order to
overcome this difficulty we consider fermion fields coupled in a nonminimal way
to torsion. Then, if the low energy action for torsion breaks parity, the left
- right asymmetry appears in the effective four - fermion interactions. That's
why we arrange all SM fermions in the left - handed components of the Dirac
spinors while the additional fermions (called technifermions in an analogy with
TC) are arranged in right - handed components of the spinors.

Once the parity breaking is admitted in the torsion action, under natural
assumptions this action has the form that leads to appearance of the
considerable asymmetry between left-handed and right-handed fermions. Due to
this asymmetry the four fermion interactions between the SM fermions are
negligible compared to that of the technifermions. That's why the four -
fermion interactions provide condensation of the technifermions while do not
affect the dynamics of the SM particles. As a result the Electroweak symmetry
is broken while the SM fermions remain massless. Some other physics is to be
added now in order to provide the appearance of their masses.

The arrangement of SM fermions and the technifermions in the left - handed and
right - handed components of the Dirac spinors allows us to introduce either
the mass term for those spinors or the interaction of the fermion bilinears
with real scalar field. Both these terms may be considered as perturbations
over the four - fermion interactions caused by torsion. We consider these two
possibilities as a source of the transition between SM fermions and
technifermions. In both cases it is demonstrated how the SM fermion mass terms
appear.

The paper is organized as follows. In the 2-nd section we describe how the term
in the torsion action that breaks parity leads to left - right asymmetry in the
effective four fermion interactions. In the 3-rd section we introduce two
spinors nonminimally coupled to torsion. The left - handed components of these
spinors are used to arrange both left - handed and right - handed SM fermions
while right - handed components of these spinors are used to arrange
technifermions. We demonstrate how the resulting four - fermion action can be
written in terms of $4$ - component SM fermions and technifermions. In the $4$
- th section we consider the whole set of the SM fermions and technifermions
and describe  how the resulting four - fermion interaction that has the form of
the Extended NJL model works in order to provide condensation of
technifermions. In section $5$ we describe how the additional real scalar field
coupled to fermions provides coupling of the SM fermions to Technifermion
condensate. In section $6$ the appearance of SM fermion masses due to an
additional scalar field is described. In section $7$ we consider how mass term
for original spinors that contain SM fermions as their left - handed components
leads to formation of SM fermion masses. In section $8$ we end with the
conclusions.

\section{Left - right asymmetry due to torsion}

The action of a massless Dirac spinor coupled nonminimally to torsion has the
form \cite{Shapiro,Shapiro_,Shapiro__}:

\begin{eqnarray}
S_f & = & \frac{1}{2}\int \{i\bar{\psi} \gamma^{\mu} \partial_{\mu} \psi -
i[\partial_{\mu}\bar{\psi}] \gamma^{\mu}\psi - \frac{1}{4}\bar{\psi}\gamma^i
(\gamma^5  \eta S_{i} + \hat{\eta}T_i) \psi\} d^4 x \label{Sf}
\end{eqnarray}

Here we assume that due to the gravitational action the inverse vierbein
$E^a_{\mu}$ is close to $\delta^a_{\mu}$ at the considered energies while usual
Christoffel symbols vanish. Axial vector torsion is $S_i =
\epsilon_{ijkl}T^{jkl}$ while vector torsion is $T_i = T_{ij}^j$. Here
$T^i_{jk}$ is usual torsion; $\eta$ and $\hat{\eta}$ are coupling constants.
Below we assume for simplicity $\eta = \hat{\eta} = 1$. (Actually, one can
always rescale $S$ and $T$ in order to make $\eta$ and $\hat{\eta}$ equal to
$1$.)

The most general quadratic in first derivatives (of $E$)  action for vector and
axial torsion (tensorial torsion is not considered) has the form:

\begin{eqnarray}
S_t &=&   \frac{1}{4}\{M_{SS}^2 \int  S^i S_i d^4x +M_{TT}^2 \int T^i T_i d^4x
- M_{ST}^2 \int S^iT_i d^4x \nonumber\\&&- M_{TS}^2 \int S^iT_i d^4x\}
\label{Stpm_}
\end{eqnarray}
Here parameters $M_{ST}=M_{TS}, M_{TT}, M_{SS}$ are of dimension of mass. We
assume all these parameters are of the same order. So, it is natural to suppose
that $|M^2_{SS}+M_{TT}^2+ M^2_{ST}+M^2_{TS}|
>> |M^2_{SS}+M_{TT}^2- M^2_{ST}-M^2_{TS}|>>|M^2_{SS} - M^2_{TT}|$. With this choice the
action can be rewritten in the form:
\begin{eqnarray}
S_t &=&  - \frac{M_+^2}{4} \int  (S^i+T^i) (S_i+T_i) d^4x - \frac{M_-^2}{4}
\int (S^i-T^i) (S_i-T_i) d^4x  \nonumber\\ && + \frac{\Delta M^2}{2} \int
(S^i+T^i)(S_i-T_i) d^4x, \label{Stpm}
\end{eqnarray}
where $M^2_-=-\frac{M^2_{SS}+M_{TT}^2+ M^2_{ST}+M^2_{TS}}{4} ,
M^2_+=\frac{-M^2_{SS}-M_{TT}^2+ M^2_{ST}+M^2_{TS}}{4} , \Delta M^2 =
\frac{M^2_{TT}-M^2_{SS}}{4}$. We also need $M^2_+ >0$. Below we assume for
simplicity $\Delta M^2 = 0$. (Actually, the consideration of the action with
$\Delta M^2 \ne 0$ can be done easily. However, it does not give anything new:
the additional four - fermion term will appear, that does not change
qualitative results provided that $M_-
>> M_+ >>\Delta M$.)

With the aid of (\ref{Stpm}) torsion in (\ref{Sf}) can be integrated out. The
resulting effective action for the fermion field is:
\begin{eqnarray}
S_f & = & \frac{1}{2}\int \{i\bar{\psi} \gamma^{\mu} \partial_{\mu} \psi -
i[\partial_{\mu}\bar{\psi}] \gamma^{\mu}\psi +
\frac{1}{32M_+^2}(\bar{\psi}_+\gamma^i \psi_+) (\bar{\psi}_+\gamma_i
\psi_+)\nonumber\\&&+ \frac{1}{32M_-^2}(\bar{\psi}_-\gamma_i \psi_-)
(\bar{\psi}_-\gamma^i \psi_-)\} d^4 x \label{Sf_}
\end{eqnarray}

Here left- and right-handed components of $\psi$ are denoted by $\psi_-$ and
$\psi_+$ respectively. We can see that if $M_+ \ne M_-$ the effective fermion
action has the left-right asymmetry.

\section{Condensation of composite Dirac field}

Let us now consider a  more complicated situation when two Dirac spinors $\psi$
and $\phi$ are coupled to torsion. We consider the fermion action of the form:
\begin{eqnarray}
S_f & = & \frac{1}{2}\int \{i\bar{\psi} \gamma^{\mu} \partial_{\mu} \psi -
i[\partial_{\mu}\bar{\psi}] \gamma^{\mu}\psi - \frac{1}{4}\bar{\psi}\gamma^i
(\gamma^5   S_{i} + T_i) \psi\} d^4 x \nonumber\\&& +\frac{1}{2}\int
\{i\bar{\phi^c} \gamma^{\mu} \partial_{\mu} \phi^c -
i[\partial_{\mu}\bar{\phi^c}] \gamma^{\mu}\phi^c -
\frac{1}{4}\bar{\phi^c}\gamma^i (\gamma^5   S_{i} + T_i) \phi^c\} d^4 x
\label{Sf2}
\end{eqnarray}

Here $\phi^c = i \gamma^2  \left(\begin{array}{c}\phi_-\\
\phi_+\end{array}\right)^*=\left(\begin{array}{c}i\sigma^2 \phi^*_+\\
-i\sigma^2 \phi^*_-\end{array}\right)$. Below we use the following
representation of
$\gamma$ matrices: $\gamma^{\mu} = \left(\begin{array}{cc}0&\sigma^{\mu}\\
\bar{\sigma}^{\mu}&0\end{array}\right)$, where $\bar{\sigma}^0 = \sigma^0 = 1;
\bar{\sigma}^i = -\sigma^i \, (i=1,2,3)$; $\gamma^5 = \left(\begin{array}{cc}1&0\\
0&-1\end{array}\right)$.

Integration over torsion leads to
\begin{eqnarray}
&&S_f  =  \int \{i\psi_+^+ {\sigma}^{\mu} \partial_{\mu} \psi_+  +i\psi_-^+
\bar{\sigma}^{\mu}
\partial_{\mu} \psi_-
+ i\phi_+^+ {\sigma}^{\mu}
\partial_{\mu} \phi_+  +
i\phi_-^+ \bar{\sigma}^{\mu} \partial_{\mu} \phi_-  \nonumber\\&&
+\frac{1}{64M_+^2}({\psi}^+_+{\sigma}^i \psi_+-{\phi}^+_-\bar{\sigma}^i
\phi_-)^2+ \frac{1}{64M_-^2}({\phi}^+_+{\sigma}^i
\phi_+-{\psi}^+_-\bar{\sigma}^i \psi_-)^2 \} d^4 x \label{Sf_2}
\end{eqnarray}

Now let us compose new spinors $\psi_t = \left(\begin{array}{c}\phi_-\\
\psi_+\end{array}\right)$ and $\psi_s = \gamma^5\left(\begin{array}{c}\psi_-\\
\phi_+\end{array}\right)$. Then we come to the following  expression for the
effective action:

\begin{eqnarray}
S_f & = & \int \{i\bar{\psi}_s \gamma^{\mu} \partial_{\mu} \psi_s +
\frac{1}{64M_-^2}(\bar{\psi}_s\gamma^i \gamma^5  \psi_s)(\bar{\psi}_s\gamma_i
\gamma^5  \psi_s)\} d^4 x \nonumber\\&&+ \int \{i\bar{\psi}_t \gamma^{\mu}
\partial_{\mu} \psi_t  +
\frac{1}{64M_+^2}(\bar{\psi}_t\gamma^i \gamma^5 \psi_t)(\bar{\psi}_t\gamma_i
\gamma^5  \psi_t)\} d^4 x \label{Sf22}
\end{eqnarray}

In this form the action both for $\psi_s$ and $\psi_t$ has the form of the
action of Nambu - Jona - Lasinio effective model of chiral symmetry breaking.
It is worth mentioning that the model with the action (\ref{Sf22}) is
nonrenormalizable and should be considered as a finite cutoff model with the
finite cutoff $\Lambda_{\chi}$. There exists the critical value of mass $M_C$
(that depends on the mentioned cutoff) such that at $M_{+} < M_C \,(M_-<M_C)$
the field $\psi_t$ ($\psi_s$) is condensed while for $M_{+} > M_C \,(M_- >M_C)$
it is not condensed.

At this point we suppose that $M_- >> M_C > M_+$. Therefore, the field $\psi_t$
is condensed while the field $\psi_s$ is not. The value of $<\bar{\psi}_t
\psi_t>$ is expected to be around $-\Lambda_T^3$ while the dynamical mass of
$\psi_t$ is about $\Lambda_T$.

Physically the parameter $\Lambda_T$ is hidden within the theory of dynamical
torsion. The scale of this theory is expected to be around $\Lambda_{\chi}$.
However, we do not see any indication that $\Lambda_T$ must be of the same
order as $\Lambda_{\chi}$. Instead we expect $\Lambda_{\chi} \sim 10 \Lambda_T$
or $\Lambda_{\chi} \sim 100 \Lambda_T$. This is in accordance with the next
section, where it is shown that in leading approximation of the NJL model the
critical value of mass is $M_C \sim 0.1 \Lambda_{\chi}$.

\section{Electroweak symmetry breaking due to torsion}

Now we are in a position to describe how torsion may provide the Electroweak
symmetry breaking of the Standard Model. Let us arrange all left - handed
fermions and right - handed fermions of the Standard Model in the left - handed
parts of the Dirac spinors. Correspondingly, the additional fields are arranged
within the right-handed parts of the given spinors. We call the mentioned
additional fermion fields technifermions. The effective action of the model at
energies much less than $\Lambda_T$ has the form:
\begin{eqnarray}
&&S_f  =  \int \{i\bar{\psi}^a_s \gamma^{\mu} D_{\mu} \psi^a_s +
\frac{1}{64M_-^2}(\bar{\psi}^a_s\gamma^i \gamma^5
\psi^a_s)(\bar{\psi}^b_s\gamma_i \gamma^5  \psi^b_s)\} d^4 x \nonumber\\&&
+\int \{i\bar{\psi}^a_t \gamma^{\mu} D_{\mu} \psi^a_t  +
\frac{1}{64M_+^2}(\bar{\psi}^a_t\gamma^i \gamma^5
\psi^a_t)(\bar{\psi}^b_t\gamma_i \gamma^5  \psi^b_t)\} d^4 x \label{Sf22s}
\end{eqnarray}
Here indices $a,b$ enumerate the mentioned Dirac spinors while the derivative
$D$ contains all Standard Model gauge fields. In a complete analogy with the
previous section we obtain condensation of technifermions $\psi_t$ provided
that $M_-
> M_C > M_+$. In the absence of the Standard Model (SM) gauge fields the $SU({\cal N})_L\otimes SU({\cal N})_R$ symmetry of
(\ref{Sf22s}) is broken down to $SU({\cal N})_V$ (here ${\cal N}$ is the total
number of SM fermions). The SM interactions act as a perturbation.

Action (\ref{Sf22s}) (except for the term with $M_-$) has the form of the
effective action for the $SU(N_g)$ Farhi - Susskind model \cite{FS} (provided
that there exist $N_g$ generations). That's why in an analogy with this
technicolor model we expect $\Lambda_T$ to be at a TeV scale.
 At the same time usual fermions $\psi_s$ remain massless and some other
 physics should be added in order to provide appearance of their observed
 masses.

 In order to make the connection with Technicolor model more explicit let us
 apply Fierz transformation to the four fermion term of (\ref{Sf22s}) for
 technifermions:
\begin{eqnarray}
S_{4,t}  &=&  \frac{1}{64M_+^2}\int (\bar{\psi}^a_t\gamma^i \gamma^5
\psi^a_t)(\bar{\psi}^b_t\gamma_i \gamma^5  \psi^b_t) d^4 x \nonumber\\
&=& \frac{1}{64M_+^2}\int\{4
(\bar{\psi}^a_{t,L}\psi^b_{t,R})(\bar{\psi}^b_{t,R} \psi^a_{t,L})\nonumber\\&&
+[(\bar{\psi}^a_{t,L}\gamma_i\psi^b_{t,L})(\bar{\psi}^b_{t,L}\gamma^i
\psi^a_{t,L})+(L\leftarrow \rightarrow R)]\} d^4 x
\end{eqnarray}
In this form the action has exactly the form of the extended NJL model for QCD
(see Eq. (4), Eq. (5), Eq. (6) of \cite{ENJL}) (with negative $G_V$, though),
where the total number of technifermions plays the role of $N_c$. So, we have
$N_c = {\cal N} = 24$; $G_S = \frac{3\Lambda^2_{\chi}}{16\pi^2M^2_+}$; $G_V =-
\frac{3\Lambda^2_{\chi}}{64\pi^2M^2_+}$. Here $\Lambda_{\chi}$ is the cutoff
that is now the physical parameter of the model. Its value depends on the
details of physics that provides the appearance of the four - fermion
interactions. In our case $\Lambda_{\chi}$ is to be calculated within the
theory of dynamical torsion.

Next, the auxiliary fields $M$, $L_i$, and $R_i$ are introduced and the new
action has the form:
\begin{eqnarray}
S  &=& \int\{ -(\bar{\psi}^a_{t,L}M^+_{ab} \psi^b_{t,R} + (h.c.)) - 16M_+^2
{\rm Tr}\, M^+M\}d^4x \nonumber\\&& +\int\{(\bar{\psi}^a_{t,L}\gamma^i
L^{ab}_i\psi^b_{t,L}) -16 M_+^2 {\rm Tr}\,L^iL_i +(L\leftarrow \rightarrow R)\}
d^4 x
\end{eqnarray}

Integrating out fermion fields we arrive at the effective action for the
mentioned auxiliary fields (and the source currents for fermion bilinears). The
resulting effective action receives its minimum at $M = m_t {\bf 1}$, where
$m_t$ plays the role of the technifermion mass (equal for all technifermions).
In leading approximation the condensate of $\psi_t$ is expressed through $m_t$
as follows:
\begin{equation}
<\bar{\psi}_t\psi_t> = -\frac{N_c}{16\pi^2}4m_t^3
\Gamma(-1,\frac{m_t^2}{\Lambda_{\chi}^2})\label{condens}
\end{equation}

Here $\Gamma(n,x) = \int_x^{\infty}\frac{dz}{z}e^{-z}z^{n}$. The gap equation
reads:

\begin{equation}
\frac{1}{G_S} =
\{\exp(-\frac{m_t^2}{\Lambda_{\chi}^2})-\frac{m_t^2}{\Lambda_{\chi}^2}
 \Gamma(0,\frac{m_t^2}{\Lambda_{\chi}^2})\}
\end{equation}
It does not depend on $G_V$. Obviously, there exists the critical value of
$G_S$: at $G_S > 1$ the gap equation has the nonzero solution for $m_t$ while
for $G_S < 1$ it has not. This means that in this approximation the
condensation of technifermions occurs at $M_+ <
\sqrt{\frac{3}{16\pi^2}}\Lambda_{\chi}\sim 0.1 \Lambda_{\chi} $.

In the absence of SM interactions the relative orientation of the SM gauge
group $G_W = SU(3)\otimes SU(2)\otimes U(1)$ and $SU({\cal N})_V$ from
$SU({\cal N})_L\otimes SU({\cal N})_R \rightarrow SU({\cal N})_V$ is
irrelevant. However, when the SM interactions are turned on, the effective
potential due to exchange by SM gauge bosons depends on this relative
orientation. Minimum of the potential is achieved in the true vacuum state and
defines the pattern of the breakdown of $G_W$. This process is known as vacuum
alignment (see, for example, \cite{Align, Align1}). The effective potential is
\cite{Align}:
\begin{eqnarray}
V(U) &=& 4 \sum_{\alpha = SU(3), SU(2), U(1); \, k} e_{\alpha}^2 \, {\rm Tr} \,
(\theta^{\alpha, k}_L U \theta^{\alpha, k}_R U^+) \,\nonumber\\&&
(-\frac{i}{2}) \int d^4 x \Delta^{\mu \nu} (x) <0|T[ J^A_{\mu L} J^A_{\nu R}
|0>\nonumber\\&& = -\frac{3}{32\pi^2} (F^2 \Delta^2)\sum_{\alpha = SU(3),
SU(2), U(1); \, k}e_{\alpha}^2 \, {\rm Tr} \, (\theta^{\alpha, k}_L U
\theta^{\alpha, k}_R U^+)
\end{eqnarray}
There is no sum over $A$ here. $\theta^{\alpha, k}_{L,R}$ are generators of
$G_W$, $\Delta^{\mu \nu} (x)$ is the gauge boson propagator, $J^A_{\mu
L;R}=(\bar{\psi}^a_{t,L;R}\lambda_{ab}^A\gamma_i\psi^b_{t,L;R})$ are
technifermion currents; matrices $\lambda_{ab}^A$ are generators of  $ SU({\cal
N})$. $U \in SU({\cal N})$ defines relative orientation of $ SU({\cal N})_V$
and $G_W$. $F$ - is the technipion constant. In general case $\Delta^2$  may be
negative. However, in \cite{Align} arguments are given in favor of
$\Delta^2>0$. Namely, it was shown that if the technicolor interactions are
renormalizable and asymptotic free, then the spectral function sum rules take
place. Then under assumption that in the spectral functions correspondent to
vector and axial vector channels of $<0|T[ J^A_{\mu L} J^A_{\nu R} |0>$ single
intermediate states dominate, one finds $\Delta^2>0$. In our case dynamical
torsion  plays the role of the technicolor interactions. That's why we need
some suppositions about the dynamical torsion theory. In particular, if we
require that this theory is renormalizable and asymptotic free (as it should in
order to be self - consistent) and that two intermediate states dominate in the
mentioned above correlator, we also have $\Delta^2>0$. Under this supposition
in a way similar to that of \cite{Align} we come to the conclusion that $G_W$
is broken in a minimal way. This means that the subgroups of $G_W$ are not
broken unless they should. The form of the condensate (\ref{condens}) requires
that $SU(2)$ and $U(1)$ subgroups are broken. That's why in a complete analogy
with $SU(N_{TC})$ Farhi - Susskind model Electroweak group in our case is
broken correctly while $SU(3)$ group remains unbroken.

\section{Transition between the left-handed and the right - handed spinors due to the scalar field}

Now let us again consider the model with two spinors $\psi$ and $\phi$ and the
following action:
\begin{eqnarray}
S_f & = & \frac{1}{2}\int \{i\bar{\psi} \gamma^{\mu} \partial_{\mu} \psi -
i[\partial_{\mu}\bar{\psi}] \gamma^{\mu}\psi - \frac{1}{4}\bar{\psi}\gamma^i
(\gamma^5   S_{i} + T_i) \psi\} d^4 x \nonumber\\&& +\frac{1}{2}\int
\{i\bar{\phi^c} \gamma^{\mu} \partial_{\mu} \phi^c -
i[\partial_{\mu}\bar{\phi^c}] \gamma^{\mu}\phi^c -
\frac{1}{4}\bar{\phi^c}\gamma^i (\gamma^5   S_{i} + T_i) \phi^c\} d^4 x
\nonumber\\&& + \int(\bar{\psi}  \psi + \bar{\phi^c}  \phi^c) H  d^4 x
\label{Sf2_}
\end{eqnarray}
Here we have introduced real scalar field $H$ with the action
\begin{eqnarray}
S_h &=&  -\frac{M_H^2}{4} \int  H^2 d^4x d^4x \label{SH}
\end{eqnarray}

After integration over $H$ and torsion we arrive at
\begin{eqnarray}
S_f & = & \int \{i\bar{\psi}_s \gamma^{\mu} \partial_{\mu} \psi_s +
\frac{1}{64M_-^2}(\bar{\psi}_s\gamma^i \gamma^5  \psi_s)(\bar{\psi}_s\gamma_i
\gamma^5  \psi_s)\} d^4 x \nonumber\\&&+ \int \{i\bar{\psi}_t \gamma^{\mu}
\partial_{\mu} \psi_t  +
\frac{1}{64M_+^2}(\bar{\psi}_t\gamma^i \gamma^5 \psi_t)(\bar{\psi}_t\gamma_i
\gamma^5  \psi_t)\} d^4 x \nonumber\\&& +\frac{1}{M_H^2}\int
(\bar{\psi}_t\gamma^5 \psi_s - \bar{\psi}_s \gamma^5\psi_t)^2 d^4 x
\label{Sf22_}
\end{eqnarray}

After Fierz rearrangement we obtain:
\begin{eqnarray}
S_f & = & \int \{i\bar{\psi}_s \gamma^{\mu} \partial_{\mu} \psi_s +
\frac{1}{64M_-^2}(\bar{\psi}_s\gamma^i \gamma^5  \psi_s)(\bar{\psi}_s\gamma_i
\gamma^5  \psi_s)\} d^4 x \nonumber\\&& + \int \{i\bar{\psi}_t \gamma^{\mu}
\partial_{\mu} \psi_t  +
\frac{1}{64M_+^2}(\bar{\psi}_t \gamma^i \gamma^5 \psi_t)(\bar{\psi}_t \gamma_i
\gamma^5  \psi_t)\} d^4 x \nonumber\\&& -
\frac{1}{M_H^2}\int\{-(\bar{\psi}_t\gamma^5 \psi_s)^2 - (\bar{\psi}_s \gamma^5
\psi_t)^2 + \frac{1}{2}[-(\bar{\psi}_s \psi_s)(\bar{\psi}_{t} \psi_t)
\nonumber\\&&+(\bar{\psi}_s \gamma^{\mu} \psi_s)(\bar{\psi}_{t} \gamma_{\mu}
\psi_t)  + \frac{1}{2}(\bar{\psi}_s
\gamma^{[\mu}\gamma^{\nu]}\psi_s)(\bar{\psi}_{t} \gamma_{[\mu}
\gamma_{\nu]}\psi_t) \nonumber\\&& - (\bar{\psi}_s
\gamma^{\mu}\gamma^{5}\psi_s)(\bar{\psi}_{t} \gamma_{\mu} \gamma^{5}\psi_t)
 -(\bar{\psi}_s
\gamma^{5}\psi_s)(\bar{\psi}_{t}  \gamma^{5}\psi_t) ] \}d^4 x \label{Sf22__}
\end{eqnarray}

Now if $\psi_t$ is condensed, $\psi_s$ acquires mass
\begin{equation}
m_s = -\frac{1}{2M_H^2}<\bar{\psi}_t \psi_t>
\end{equation}

\section{Masses of Standard Model fermions}

Now we are ready to describe how Standard Model fermion masses appear. Let us
consider the fermion action in the form:
\begin{eqnarray}
S_f & = &  \int \{i\bar{\psi}_a \gamma^{\mu} D_{\mu} \psi_a -
\frac{1}{8}\bar{\psi}_a\gamma^i (\gamma^5   S_{i} + T_i) \psi_a\} d^4 x
\nonumber\\&& +\int \{i\bar{\phi^c}_b \gamma^{\mu} \bar{D}_{\mu} \phi_b^c -
\frac{1}{8}\bar{\phi^c}_b\gamma^i (\gamma^5   S_{i} + T_i) \phi_b^c\} d^4 x
\nonumber\\&& + \int(\delta_{a a^{\prime}}\bar{\psi}_{a} \psi_{a^{\prime}}
+\eta_{b b^{\prime}}\bar{\phi}_{b} \phi_{b^{\prime}}) H d^4 x \label{Sf2SM}
\end{eqnarray}

Here index $a$ enumerates left-handed SM fermions while $b$ enumerates
right-handed SM fermions.  $\eta$ is hermitian matrix of couplings. Integrating
out $T$, $S$, and $H$, we obtain:
\begin{eqnarray}
S_f & = & \int \{i\bar{\psi}^a_s \gamma^{\mu} D_{\mu} \psi^a_s +
\frac{1}{64M_-^2}(\bar{\psi}^a_s\gamma^i \gamma^5
\psi^a_s)(\bar{\psi}^b_s\gamma_i \gamma^5  \psi^b_s)\} d^4 x \nonumber\\&&+
\int \{i\bar{\psi}^a_t \gamma^{\mu} D_{\mu} \psi^a_t  +
\frac{1}{64M_+^2}(\bar{\psi}^a_t\gamma^i \gamma^5
\psi^a_t)(\bar{\psi}^a_t\gamma_i \gamma^5 \psi^a_t)
\nonumber\\&&-\frac{1}{M_H^2}(\delta_{a a^{\prime}}\bar{\psi}_{-,a}
\psi_{+,a^{\prime}} + \delta_{a a^{\prime}}\bar{\psi}_{+,a}
\psi_{-,a^{\prime}}\nonumber\\&& +\eta_{b b^{\prime}}\bar{\phi}_{-,b}
\phi_{+,b^{\prime}}+\eta_{b b^{\prime}}\bar{\phi}_{+,b} \phi_{-,b^{\prime}})^2
d^4 x \label{Sf22SM_}
\end{eqnarray}
Here we have introduced the usual SM Dirac fermions
$\psi^a_s = \gamma^5 \left(\begin{array}{c}\psi^a_-\\
\phi^a_+\end{array}\right)$ and technifermions $\psi^a_t = \left(\begin{array}{c}\phi^a_-\\
\psi^a_+\end{array}\right)$.

Let us now suppose that due to torsion technifermions are condensed. Vacuum
alignment due to SM interactions was discussed in Section 4. Now the vacuum
alignment should take into account mass term of (\ref{Sf2SM}) as well. Namely,
this term also plays the role of perturbation that influences the alignment of
vacuum. Rather obvious, however, that if $\eta$ is diagonal in $SU(2)$ and
$SU(3)$ indices and does not depend on color, then the perturbation of this
type does not destroy the correct picture of Electroweak symmetry breaking.
Then the usual condensate appears:

\begin{equation}
<\bar{\psi}_t^a \psi^b_t> = - \delta^{ab} \Lambda^3_{T}
\end{equation}

After Fierzing we obtain the mass matrix for the SM fermions:
\begin{equation}
m_s = \frac{\Lambda_T^3}{2M_H^2}\eta
\end{equation}

\section{Mass term for $\psi$ and $\phi$}

In this section we consider another possibility to give masses to the SM
fermions. Namely, let us consider action (\ref{Sf2SM}) with the additional mass
 terms for spinors $\psi$ and $\phi$:

\begin{eqnarray}
S_f & = & \int \{i\bar{\psi}_a \gamma^{\mu} D_{\mu} \psi_a -
\frac{1}{8}\bar{\psi}_a\gamma^i (\gamma^5   S_{i} + T_i) \psi_a\} d^4 x
\nonumber\\&& +\int \{i\bar{\phi^c}_b \gamma^{\mu} \bar{D}_{\mu} \phi_b^c -
\frac{1}{8}\bar{\phi^c}_b\gamma^i (\gamma^5   S_{i} + T_i) \phi_b^c\} d^4 x
\nonumber\\&& - \int(\delta_{a a^{\prime}}\bar{\psi}_{a} \psi_{a^{\prime}}
+\eta_{b b^{\prime}}\bar{\phi}_{b} \phi_{b^{\prime}}) m_0 d^4 x \label{Sf2SM__}
\end{eqnarray}
Here $m_0$ is the constant of the dimension of mass while $\eta$ is the
hermitian matrix of couplings. As in the previous section we imply here that
$\eta$ is diagonal in $SU(2)$ and $SU(3)$ indices and does not depend on color.
Taken in such form the mass term of (\ref{Sf2SM__}) considered as a
perturbation does not destroy the correct vacuum alignment. Integrating over
torsion we obtain:
\begin{eqnarray}
S_f & = & \int \{i\bar{\psi}^a_s \gamma^{\mu} D_{\mu} \psi^a_s +
\frac{1}{64M_-^2}(\bar{\psi}^a_s\gamma^i \gamma^5
\psi^a_s)(\bar{\psi}^b_s\gamma_i \gamma^5  \psi^b_s)\} d^4 x \nonumber\\&&+
\int \{i\bar{\psi}^a_t \gamma^{\mu} D_{\mu} \psi^a_t  +
\frac{1}{64M_+^2}(\bar{\psi}^a_t\gamma^i \gamma^5
\psi^a_t)(\bar{\psi}^b_t\gamma_i \gamma^5  \psi^b_t)\} d^4 x \nonumber\\&& -
\int(\delta_{a a^{\prime}}\bar{\psi}_{-,a} \psi_{+,a^{\prime}} + \delta_{a
a^{\prime}}\bar{\psi}_{+,a} \psi_{-,a^{\prime}}\nonumber\\&& +\eta_{b
b^{\prime}}\bar{\phi}_{-,b} \phi_{+,b^{\prime}}+\eta_{b
b^{\prime}}\bar{\phi}_{+,b} \phi_{-,b^{\prime}}) m_0 d^4 x \label{Sf22SM___}
\end{eqnarray}

Next, we neglect the terms with $M_-$ and SM gauge fields that are to be
considered as perturbations. We also introduce the auxiliary fields as in ENJL
approach:
\begin{eqnarray}
S_f & = & \int \{i\bar{\psi}^a_s \gamma^{\mu} \partial_{\mu} \psi^a_s+
i\bar{\psi}^a_t \gamma^{\mu} \partial_{\mu} \psi^a_t\}d^4x \nonumber\\&&\int\{
-(\bar{\psi}^a_{t,L}M^+_{ab} \psi^b_{t,R} + (h.c.)) - 32M_+^2 {\rm Tr}\,
M^+M\}d^4x \nonumber\\&& +\int\{(\bar{\psi}^a_{t,L}\gamma^i
L^{ab}_i\psi^b_{t,L}) -16 M_+^2 {\rm Tr}\,L^iL_i +(L\leftarrow \rightarrow R)\}
d^4 x\nonumber\\&& - \int(\delta_{a a^{\prime}}\bar{\psi}_{-,a}
\psi_{+,a^{\prime}} + \delta_{a a^{\prime}}\bar{\psi}_{+,a}
\psi_{-,a^{\prime}}\nonumber\\&& +\eta_{b b^{\prime}}\bar{\phi}_{-,b}
\phi_{+,b^{\prime}}+\eta_{b b^{\prime}}\bar{\phi}_{+,b} \phi_{-,b^{\prime}})
m_0 d^4 x \label{Sf22SMNJL}
\end{eqnarray}

 Integration over technifermions leads to appearance of the effective potential for $M$ that has its minimum at
 $M = m_t {\bf 1}$. So, $M = m_t {\bf 1} + H$, where vacuum value of $H$ is zero. Thus we get:
 \begin{eqnarray}
&&S_f  =  \int \bar{\psi}_s i\gamma^{\mu} \partial_{\mu}\psi_s d^4x+
S_{eff}[L,R,H] \nonumber\\&&-m_0^2\int {\left(\begin{array}{c}\psi_{s,L}\\
-\eta \psi_{s,R}\end{array}\right)}^+\gamma^0   [i \gamma^{\mu}
D_{\mu} - m_t {\bf 1}-H]^{-1} \left(\begin{array}{c}\psi_{s,L}\\
-\eta \psi_{s,R}\end{array}\right)d^4x \label{Sf22SMNJL_}
\end{eqnarray}
Here $D_{\mu} =  \partial_{\mu} -i \frac{1+\gamma_5}{2}L_{\mu}-i
\frac{1-\gamma_5}{2}R_{\mu}$. Now our supposition is that $m_t >> m_0$. We
neglect fluctuations of $H$, $L$, and $R$  around their zero vacuum values and
arrive at:

\begin{eqnarray}
&&S_f  =  \int \bar{\psi}_s (i\gamma^{\mu} \partial_{\mu}-m_0^2 \eta [m_t
]^{-1}\psi_sd^4x \label{Sf22SMNJLF}
\end{eqnarray}

 As a result the mass term for
$\psi_s$ appears with the mass matrix

\begin{equation}
m_s = \frac{m_0^2}{m_t}\eta
\end{equation}

It is worth mentioning that in order to consider the mass terms for $\psi$ and
$\phi$ as  perturbations we need $\eta m_0 << 100$ Gev. This means that the
consideration of the present section is valid only for SM particles with masses
less than $10$ Gev. Thus, the  $t$ - quark mass generation needs an additional
consideration.

\section{Conclusions}

In the present paper we suggest the possibility that dynamical torsion (with
the scale one or two order of magnitude larger than the TeV scale ) coupled
nonminimally to fermions can provide DEWSB. In order to construct the
appropriate model we allow the parity breaking term to appear in torsion
action. As a result, left - right asymmetry appears in the effective four -
fermion interactions. We arrange all SM fermions in left - handed components of
the Dirac spinors while right - handed components are reserved for
technifermions. Due to the mentioned asymmetry the four - fermion terms that
contain SM fermions are neglibible compared to the four - fermion term that
contains technifermions. The latter term has the form of the action for
Extended NJL model. This allows us to conclude that under the given assumptions
about the torsion action the technifermions are condensed and cause the
appearance of $W$ and $Z$ - boson masses. In order to provide the correct
vacuum alignment we also made some assumptions about the dynamical torsion
theory similar to the assumptions about the technicolor interactions made in
\cite{Align}.

In order to provide appearance of masses for the SM fermions we consider two
possibilities. The first possibility is to couple fermions to the real scalar
field. Due to ejection of the scalar particle SM fermion may be transformed to
technifermion. It is implied that the scalar mass is well above $\Lambda_T$
(the scale of the technifermion condensation). As a result the SM fermions are
coupled to technifermion condensate and acquire masses. (The similar mechanism
is used in the so-called bosonic technicolor.) The second possibility is to add
directly the mass term for the Dirac spinors that contain SM fermions as their
left-handed components. This term is considered as a perturbation over the four
- fermion interactions caused by torsion. We demonstrate that in leading
approximation of the ENJL model appeared due to torsion the mass term for the
SM fermions appears. Probably, the most attractive feature of the given
constructions is that the transitions between SM fermions and technifermions
are provided while dangerous FCNC do not appear. However, in the first case the
ejection of a massive scalar is accompanied by usual Hierarchy problem as the
correspondent mass receives quadratically divergent contributions from loop
corrections. In the second case (the explicit mass term for the Dirac fermions)
the Hierarchy problem does not appear.

There is the important question about the scale of torsion mass parameters and
the mass parameter entering (\ref{Sf2SM__}) that gives rise to SM mass matrix.
Actually, if one assumes that quantum gravity theory enters the game at the
energies of the order of Planck mass $m_p$, such mass parameters might be
generated dynamically and, therefore, receive values at a $m_p$ scale.
Therefore in order to make the considerations of the present paper self -
consistent we must suppose that  there exists a mechanism within the $m_p$
scale theory that forbids dynamical generation of torsion mass as well as $m_0$
from (\ref{Sf2SM__}). Actually, we may suppose that there is no quantum theory
of Riemannian geometry at all. Then the dynamical torsion theory may be thought
of as a gauge theory of Lorentz group that is defined in Minkowsky space
\cite{Minkowsky}. This theory may have a scale slightly above $1$ TeV. In this
approach there is no problem with the scale $m_p$ at all. As for the classical
gravity, it may appear, for example, as an entropic force \cite{Entropy_force}.

It is worth mentioning that the FCNC are absent in our approach. Therefore, the
main difficulty of ETC models is avoided. However, the technifermions
contribute to the Electroweak polarization operators and the consideration of
these contributions is necessary in order to understand is our approach
realistic or not. However, we consider this issue to be out of the scope of the
present paper.

The main difficulty of our approach is, of course, that the Extended NJL model
that appears after the integration over torsion is not renormalizable and is to
be considered as a finite cutoff model. The results depend on the cutoff
$\Lambda_{\chi}$, that is, therefore, a physical parameter of the theory. This
parameter is hidden within the theory of dynamical torsion and is expected to
be one or two order of magnitude larger than the scale $\Lambda_T$. So, we
expect $\Lambda_{\chi} \sim 10$ TeV or $\Lambda_{\chi} \sim 100$ TeV. At the
same time the parameters of the effective low energy action for torsion must be
$|M^2_{SS}+M_{TT}^2+ M^2_{ST}+M^2_{TS}|
>> |M^2_{SS}+M_{TT}^2- M^2_{ST}-M^2_{TS}|>>|M^2_{SS} - M^2_{TT}|$. The physical mechanism for
appearance of such a Hierarchy remains unknown. Also the complete theory of
dynamical torsion is unknown that is to appear somewhere above the TeV scale.

This work was partly supported by RFBR grants  09-02-00338, 08-02-00661, by
Grant for leading scientific schools 6260.2010.2.

\clearpage


\begin{thebibliography}{99}

\bibitem{Rovelli}
A.Perez, C.Rovelli, Phys.Rev. D73 (2006) 044013,  ArXiv:gr-qc/0505081

\bibitem{Xue}
She-Sheng Xue, Phys.Lett.B665:54-57,2008, ArXiv:0804.4619

\bibitem{Alexander1}
S.Alexander, T.Biswas, G.Calcagni, Phys. Rev. D 81, 043511 (2010),
ArXiv:0906.5161

\bibitem{Alexander2}
S.Alexander, D.Vaid, ArXiv:hep-th/0609066

\bibitem{Shapiro}
A.S.Belyaev, I.L.Shapiro, Nucl.Phys. B543 (1999) 20-46, ArXiv:hep-ph/9806313

\bibitem{Shapiro_}
I.L.Shapiro, Mod.Phys.Lett.A9:729-733,1994

\bibitem{Shapiro__}
I.L.Shapiro, Phys.Rept.357:113,2002

\bibitem{Shapiro___}
V.~A.~Kostelecky, N.~Russell and J.~Tasson,
   %``New Constraints on Torsion from Lorentz Violation,''
   Phys.\ Rev.\ Lett.\  {\bf 100}, 111102 (2008)
   [arXiv:0712.4393 [gr-qc]].


\bibitem{Technicolor0}
S. Weinberg, Phys. Rev. D 19 (1979) 1277; L. Susskind, Phys. Rev. D 20 (1979)
2619.


\bibitem{Technicolor}
 Christopher T. Hill, Elizabeth H. Simmons,
 Phys.Rept. 381 (2003) 235-402;
Erratum-ibid. 390 (2004) 553-554

\bibitem{Technicolor_}
Kenneth Lane, hep-ph/0202255

\bibitem{Technicolor__}
R. Sekhar Chivukula, hep-ph/0011264

\bibitem{ENJL}
J.Bijnens, C.Bruno, E. de Rafael, Nucl.Phys. B390 (1993) 501-541, hep-ph/920623

\bibitem{ExtendedTechnicolor0}
S. Dimopoulos and L. Susskind, Nucl. Phys. B155 (1979) 237; E. Eichten and K.
Lane, Phys. Lett. B90 (1980) 125.


\bibitem{ExtendedTechnicolor}
 Thomas Appelquist, Neil Christensen, Maurizio Piai, Robert Shrock, Phys.Rev. D70 (2004) 093010

\bibitem{ExtendedTechnicolor_}
 Adam Martin, Kenneth Lane, Phys.Rev. D71 (2005) 015011

\bibitem{ExtendedTechnicolor__}
 Thomas Appelquist, Maurizio Piai, Robert Shrock, Phys.Rev. D69
(2004) 015002

\bibitem{ExtendedTechnicolor___}
Robert Shrock, hep-ph/0703050

\bibitem{ExtendedTechnicolor____}
 Adam Martin, Kenneth Lane, Phys.Rev. D71 (2005) 015011

\bibitem{walking}

Thomas Appelquist, Anuradha Ratnaweera, John Terning, L. C. R. Wijewardhana,
Phys.Rev. D58 (1998) 105017


\bibitem{minimal_walking}
R. Foadi, M.T. Frandsen, T. A. Ryttov, F. Sannino, arXiv:0706.1696

\bibitem{minimal_walking_}
Sven Bjarke Gudnason, Chris Kouvaris, Francesco Sannino,   Phys.Rev. D73 (2006)
115003

\bibitem{minimal_walking__}
D.D. Dietrich (NBI), F. Sannino (NBI), K. Tuominen, Phys.Rev. D72 (2005) 055001

\bibitem{BosonicTC}
Elizabeth H. Simmons, Nucl.Phys.B312:253,1989.

\bibitem{BosonicTC_}
Michael Dine, Alex Kagan, Stuart Samuel. Phys.Lett.B243:250-256,1990.


\bibitem{Kagan}
Alex Kagan,  CCNY-HEP-91-12,  Proc. of 15th Johns Hopkins Workshop on Current
Problems in Particle Theory, Baltimore, MD, Aug 26-28, 1991,
 Johns Hopkins Wrkshp 1991:217-242
(QCD161:J55:1991)

\bibitem{Dobrescu_Kagan}
Bogdan A. Dobrescu, Nucl.Phys.B449:462-482,1995

\bibitem{Kagan_}
D.Atwood, A.Kagan and T.G.Rizzo,
%``Constraining Anomalous Top Quark Couplings At The TeVatron,''
Phys.\ Rev.\ D {\bf 52}, 6264 (1995) [arXiv:hep-ph/9407408].
%%CITATION = PHRVA,D52,6264;%%

\bibitem{Kagan__}
A.L.Kagan,
%``Implications of TeV flavor physics for the Delta I = 1/2 rule and
%BR(l)(B),''
Phys.\ Rev.\ D {\bf 51}, 6196 (1995) [arXiv:hep-ph/9409215].
%%CITATION = PHRVA,D51,6196;%%

\bibitem{Dobrescu_Kagan_}
B.A.~Dobrescu and E.H.Simmons,
%``Top-bottom splitting in technicolor with composite scalars,''
Phys.\ Rev.\ D {\bf 59}, 015014 (1999) [arXiv:hep-ph/9807469].
%%CITATION = PHRVA,D59,015014;%%

\bibitem{Dobrescu_Kagan__}
B.A.Dobrescu and J.Terning,
%``Negative contributions to S in an effective field theory,''
Phys.\ Lett.\ B {\bf 416}, 129 (1998) [arXiv:hep-ph/9709297].
%%CITATION = PHLTA,B416,129;%%

\bibitem{Z2010_2}
M.A.Zubkov, arXiv:1003.3538




\bibitem{FS}
E.Farhi, L.Susskind, Phys.Rev.D 20, 3404, 1979











\bibitem{Align}
J. Preskill, Nucl. Phys. B177, 21 (1981)

\bibitem{Align1}
 M. E. Peskin, Nucl. Phys. B175, 197
(1980).


\bibitem{Minkowsky}
Nakia Carlevaro, Orchidea Maria Lecian, Giovanni Montani, Int. J. Mod. Phys. A
23, 1282-1285 (2008)

\bibitem{Minkowsky_}
Nakia Carlevaro, Orchidea Maria Lecian, Giovanni Montani,
Mod.Phys.Lett.A24:415-427,2009

\bibitem{Entropy_force}
 Erik P.
Verlinde, arXiv:1001.0785

\end{thebibliography}
\end{document}